# Semi-supervised Learning with Robust Loss in Brain Segmentation

**Hedong Zhang, Anand A. Joshi**

University of Southern California

**Abstract.** In this work, we used a semi-supervised learning method to train deep learning model that can segment the brain MRI images. The semi-supervised model uses less labeled data, and the performance is competitive with the supervised model with full labeled data. This framework could reduce the cost of labeling MRI images. We also introduced robust loss to reduce the noise effects of inaccurate labels generated in semi-supervised learning.

## 1 Introduction

Image segmentation is an important part of medical image processing. The rapid-developing deep neural networks (DNNs) can be used for this purpose. The purpose of our method is to segment brain MRI images using DNN. MRI images are common in medical field and have higher resolutions compared with normal images. For supervised DNNs, the model can learn how to segment image using labeled images. However, for MRI images, the cost is very expensive if all images are labeled by experts.

One approach to decrease the cost is semi-supervised learning. In this work, we split the supervise method to semi-supervise method, coupled by supervised and self-supervised processes. The self-supervised part only need data without labeled, which avoid suffering from lack of training data.

In the self-supervised part, the pseudo-labels are often mislabeled. And it is hard to promise there is no outliers in the ground truth. We used BCE robust loss to relieve the impact caused by the incorrect labels.

## 2 Method

### 2.1 Background

The semi-supervised part uses the idea of consistency regularization. By minimizing multiple predictions of the same sample with weak and strong augmentations, the model can achieve better generalization performance [1][4][3]. The model is trained with both supervised loss and unlabeled loss

$$\sum_{i=1}^{n}\sum_{b=1}^{m} \| F^b(T^w(x_i)) - F^b(T^s(x_i)) \| \tag{1}$$

where $F^b$ is the predict model, $T^w$ and $T^s$ are different data augmentation and $x_i$ is the input image.

## 2.2 Semi-Supervised Loss

We implied three kinds of augmentations: one weak and one strong augmentation. The weak augmentation in the experiment is gaussian noise. Specifically, for each pixel value, we randomly add the value to the original pixel value in a Gaussian distribution with 50% probability. The strong augmentation we used is style augmentation. It grabs the texture from the base image and mixes the style with the target image.

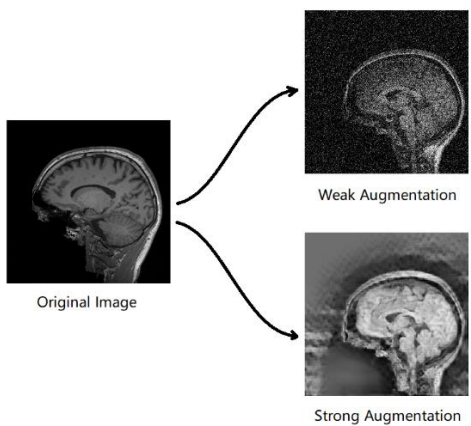

Figure 1: Weak & Strong Augmentations
Weak: *Gaussian Noise*
Strong: *Style Augmentaion*

**Pseudocode of Semi-Supervised Training**

```
# x_l: input with label, gt: groundtruth
# x_u: input without label
for (x_l, gt), x_u in loader:

    # one weak view and two strong views as input
    x_l = weak_augmentation(x_l)
    x_w = weak_augmentation(x_u)
    x_s = strong_augmentation(x_w)

    # three predictions (p_l, p_w, p_s): BxCxHxW
    p_l, p_w, p_s = model(torch.cat((x_l, x_w, x_s))).chunk(3)

    # pseudo label
    mask_w = p_w.argmax(dim=1).detach()

    # supervised loss
    loss_x = criterion(p_l, gt)
    # unsupervised loss
    loss_u = criterion(p_s, mask_w)

    loss = (loss_x + loss_u) / 2.0
```

At the beginning of training, the predict accuracy of model could be very low. We used two loss functions to filter noisy labels. First can be formulated as:

$$\mathcal{L}^u = \frac{1}{B}\sum \mathbb{I}(\max(p^w) > \tau) \, CE(p^w, p^s) \tag{2}$$

$B$ is the batch size of unlabeled images, $\tau$ is the threshold. We only calculate the loss of pixels whose probability is over the threshold $\tau$.

Another method is using robust. We used Beta Cross-Entropy (BEC) loss in training unlabeled data [5][6]. The BCE loss can be expressed as:

$$\mathcal{L}_{BCE} = \frac{\beta + 1}{\beta}(1 - p(y|x))^{\beta} + \sum_{k=1}^{K} p(k|x)^{\beta+1} \qquad (3)$$

## 3 Experiments and Results

In this section, we first provide the details of the framework. There are 9 classes: background, White Matter (WM), Gray Matter (GM), Cerebro-Spinal Fluid (CSF), bones, skin, cavities, eyes, and ventricles. We use dice scores of each tissue to evaluate performance of models and compared the performance with baseline. To show the superiority of semi-supervised loss, we also compared with model with different data augmentations.

### 3.1 Experiment details

The baseline we used is TransUNet [2]. For fair comparison, we set the same batch size, learning rate, training dataset. We modified the corresponding parts of the data augmentation and/or loss formulations in ablation experiments. The results are shown in the follow section. For the semi-supervised training, we use 50% training data as labeled images and the other 50% as unlabeled images.

### 3.2 Results

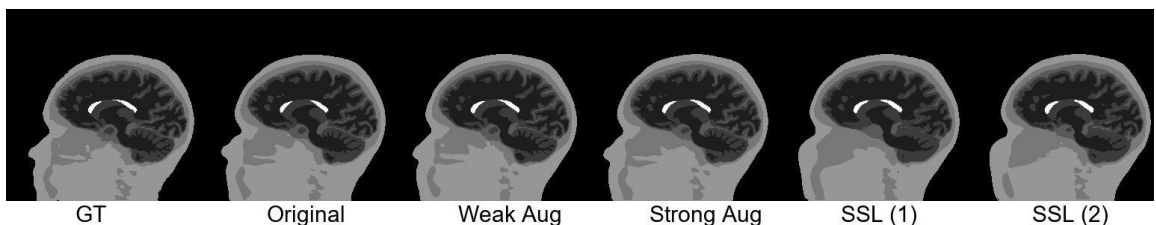

Figure 2. The output of different models

| Model | Background | WM | GM | CSF | Bones | Skin | Cavities | Eyes | Ventricles |
|---|---|---|---|---|---|---|---|---|---|
| Baseline | 0.993 | 0.886 | 0.849 | 0.750 | 0.873 | 0.926 | 0.695 | 0.731 | 0.907 |
| Weak Augmentation | **0.994** | **0.917** | 0.886 | 0.794 | 0.889 | **0.939** | 0.729 | 0.781 | **0.934** |
| Strong Augmentation | **0.994** | **0.917** | **0.887** | **0.801** | **0.890** | **0.939** | 0.774 | 0.782 | 0.933 |
| Semi-Supervise (1) | 0.992 | 0.872 | 0.848 | 0.705 | 0.841 | 0.913 | **1** | 0.827 | 0.921 |
| Semi-Supervise (2) | 0.992 | 0.866 | 0.846 | 0.693 | 0.851 | 0.913 | **1** | **1** | 0.905 |

Table 1. Weak augmentation is gaussian noise. Strong augmentation is style augmentation. Semi-Supervise (1): use cross-entropy with threshold. (2): use BCE Loss

## 4 Conclusion

We study a semi-supervised method to reduce the requirement of label training images. With using only 50% labeled data, there was no significant performance degradation in semi-supervised model, and there was a obvious performance improvement in several tissues where the original model did not perform well.